\documentclass[12pt,a4paper]{article}
\usepackage[top=25truemm,bottom=35truemm,left=25truemm,right=25truemm]{geometry}
\usepackage{amsmath,amssymb}
\usepackage[dvipdfmx]{graphicx}

\begin{document}
\setlength{\baselineskip}{18pt}
\begin{titlepage}
\begin{flushright}
\begin{tabular}{l}
%  SU-HET-06-2016\\
\end{tabular} 
\end{flushright}

\vspace*{1.2cm}
\begin{center}
{\Large\bf 
Naturalness and lepton number/flavor violation in inverse seesaw models
}
\end{center}
\lineskip .75em
\vskip 1.5cm

\begin{center}
{\large Naoyuki Haba}$^1$,
{\large Hiroyuki Ishida}$^{1,2}$, 
{\large and Yuya Yamaguchi}$^{1,3}$\\

\vspace{1cm}

$^1${\it Graduate School of Science and Engineering, Shimane University,\\
 Matsue 690-8504, Japan}\\
$^2${\it Physics Division, National Center for Theoretical Sciences,\\
 Hsinchu, Taiwan 300}\\
$^3${\it Department of Physics, Faculty of Science, Hokkaido University,\\
 Sapporo 060-0810, Japan}\\

\vspace{10mm}
{\bf Abstract}\\[5mm]
{\parbox{14.5cm}{\hspace{5mm}
%%%%%%%%%%%%%%%%%%%%%%%%%%%%%%%%%%%%%
%             ABSTRACT                      %
%%%%%%%%%%%%%%%%%%%%%%%%%%%%%%%%%%%%%

We introduce three right-handed neutrinos and three sterile neutrinos,
 and consider an inverse seesaw mechanism for neutrino mass generation.
From naturalness point of view,
 their Majorana masses should be small,
 while it induces a large neutrino Yukawa coupling.
Then, a neutrinoless double beta decay rate can be enhanced,
 and a sizable Higgs mass correction is inevitable.
We find that the enhancement rate can be more than ten times
 compared with a standard prediction from light neutrino contribution alone,
 and an analytic form of heavy neutrino contributions to the Higgs mass correction.
In addition,
 we numerically analyze the model,
 and find almost all parameter space of the model can be complementarily searched
 by future experiments of neutrinoless double beta decay and $\mu \to e$ conversion.
}}
\end{center}
\end{titlepage}

%%%%%%%%%%%%%%%%%%%%%%%%%%%%%%%%%%%%%%
\section{Introduction}
%%%%%%%%%%%%%%%%%%%%%%%%%%%%%%%%%%%%%%

The experimental results on the neutrino oscillation have established an exact evidence of neutrino masses.
Since the origin of neutrino masses cannot be explained within the standard model (SM),
 there are a lot of models towards explaining the tiny neutrino masses naturally.
The type-I seesaw model~\cite{seesaw} is one of the simplest idea,
 in which right-handed neutrinos are introduced.
There is a large parameter space for their masses,
 and they could be solve some phenomenological problems:
 short baseline neutrino oscillation anomalies with a eV mass\footnote{
The neutrino oscillation anomalies have been reported by some experiments:
 reactor~\cite{Mueller:2011nm,Huber:2011wv,Mention:2011rk},
 accelerator~\cite{Aguilar:2001ty}-\cite{Aguilar-Arevalo:2013pmq}
 and Gallium~\cite{Acero:2007su,Giunti:2010zu}.},
 relic abundance of dark matter with a keV mass,
 and baryon asymmetry of the universe (BAU) with GeV to TeV masses with a sufficient fine-tuning
 [see Ref.~\cite{Drewes:2013gca} for a review].

% The Majorana masses of right-handed neutrinos violate lepton number conservation.
% The lepton number violation, however, is must be very small by the absence of neutrinoless double beta decay~\cite{KamLAND-Zen:2016pfg}.
% This fact may suggest approximate lepton number conservation,
%  and the small Majorana masses are natural in the sense of 't Hooft~\cite{'tHooft:1979bh}.
% In the type-I seesaw model,
%  the BAU can be simply explained by leptogenesis~\cite{Fukugita:1986hr},
%  which requires the heavy neutrino mass $\gtrsim 10^9\,{\rm GeV}$.
% However, it is impossible to search such a heavy mass region.

In the conventional type-I seesaw model,
 the tiny neutrino masses and the BAU can be simultaneously explained without a fine-tuning,
 in which the right-handed neutrinos are typically heavier than $10^9\,{\rm GeV}$~\cite{Fukugita:1986hr}.
It is, however, impossible to search such heavy right-handed neutrinos directly.
For the indirect searches,
 since the Majorana masses of right-handed neutrinos violate the lepton number conservation,
 lepton number violation processes might be detectable.
Nevertheless, the absence of the neutrinoless double beta decay at the moment~\cite{KamLAND-Zen:2016pfg}
 suggests the approximate lepton number conservation.
Then, the Majorana masses should be small from the naturalness in the sense of 't Hooft~\cite{'tHooft:1979bh}.

On the other hand,
 inverse seesaw models~\cite{Wyler:1982dd,Mohapatra:1986bd,GonzalezGarcia:1988rw}
 are much interesting from experimental point of view.
In the inverse seesaw models,
 there are right-handed neutrinos which couple with left-handed neutrinos
 and sterile neutrinos which do not couple with left-handed neutrinos.
Compared with the type-I seesaw model,
 their Majorana masses can be smaller, which is preferred by the naturalness,
 and sizable neutrino Yukawa couplings are allowed.
Thus, the inverse seesaw model has been strongly constrained by
 lepton flavor violations~\cite{Atre:2009rg},
 cosmology~\cite{Drewes:2015iva}
 and collider experiments~\cite{Dev:2013wba,Das:2014jxa,Das:2015toa}.
Moreover, the inverse seesaw model can be highly testable by
 precise measurements of lepton flavor violation
 (for example,
 $\mu \to e \gamma$~\cite{Baldini:2013ke},
 $\mu \to eee$~\cite{Blondel:2013ia},
 and $\mu \to e$ conversion~\cite{Barlow:2011zza}),
 and high energy/luminosity collider experiments~\cite{Chen:2011hc}-\cite{DeRomeri:2016gum}.
In this paper,
 we focus on the (3, 3) inverse seesaw model,
 in which the number of both right-handed neutrinos and sterile neutrinos are three,
 and we do not consider the phenomenological issues mentioned above.

This paper is organized as follows.
In Sec.~\ref{sec:mass},
 we briefly review the inverse seesaw mechanism and define our setup.
In Sec.~\ref{sec:beta},
 we investigate heavy neutrino contributions for the neutrinoless double beta decay,
 and find an analytic form of the heavy neutrino contribution to the effective neutrino mass.
In Sec.~\ref{sec:mass_corr},
 we also investigate the Higgs mass correction coming from heavy neutrinos,
 which should be not so larger than the electroweak (EW) scale from the naturalness point of view.
In Sec.~\ref{sec:numerical},
 we show numerical results of heavy neutrino contributions to 
 the effective neutrino mass and the Higgs mass correction.
Finally, we summarize our results in Sec.~\ref{sec:summary}.

% This paper is organized as follows.
% In Sec.~\ref{sec:mass},
%  we briefly review the inverse seesaw mechanism and define our setup.
% In Sec.~\ref{sec:beta},
%  we investigate heavy neutrino contributions for the neutrinoless double beta decay.
% Its rate is proportional to the effective neutrino mass,
%  and it is useful to estimate a contribution from heavy neutrinos.
% We find an analytic form of the heavy neutrino contribution to the effective neutrino mass.
% It is strongly suppressed for a large heavy neutrino mass $\gtrsim {\rm GeV}$,
%  while can be much larger than the active neutrino contribution in $\sim 0.1\,{\rm GeV}$ region.
% In Sec.~\ref{sec:mass_corr},
%  we investigate the Higgs mass correction coming from heavy neutrinos,
%  which should be not so larger than the electroweak (EW) scale from the naturalness point of view.
% We find the minimal value of Higgs mass correction is determined for a given heavy neutrino mass,
%  and it is usually larger than the type-I seesaw case.
% In Sec.~\ref{sec:numerical},
%  we show numerical results of heavy neutrino contributions to 
%  the effective neutrino mass and the Higgs mass correction.
% Then, we find that almost all parameter space of the inverse seesaw model can be searched
%  by future experiments of $\mu \to e$ conversion.
% Finally, we summarize our results in Sec.~\ref{sec:summary}.

% \newpage
%%%%%%%%%%%%%%%%%%%%%%%%%%%%%%%%%%%%%%
\section{Mass spectrum in inverse seesaw models} \label{sec:mass}

In this section, we explain a mass spectrum in inverse seesaw models.
To realize the inverse seesaw mechanism,
 we introduce three pairs of SM gauge singlet fields ($N$, $S$),
 which have the lepton number $L=1$.
Then, the Lagrangian contains
\begin{eqnarray}
	\mathcal{L} = i \overline{N} \gamma^\mu \partial_\mu N + i \overline{S} \gamma^\mu \partial_\mu S
		- M_N \overline{N} S
		- \frac{\mu_N}{2} \overline{N^c}N - \frac{\mu_S}{2} \overline{S^c}S
		- Y_\nu \overline{N}\tilde{\Phi} \ell + {\rm h.c.}\,,
\label{Lag}
\end{eqnarray}
 where $\ell$ is the lepton doublet, which has the lepton number $L=1$,
 and $\tilde{\Phi}=i\sigma_2 \Phi$ is the charge conjugation of the Higgs doublet $\Phi$.
The $3\times 3$ complex matrices $M_N$, $\mu_{N,S}$, and $Y_{\nu}$ stand for
 Dirac-type mass, Majorana masses, and neutrino Yukawa coupling, respectively.
We have assumed that $S$ does not couple with $\ell$ at tree level,\footnote{
In Ref.~\cite{Law:2013gma},
 the authors introduced vector-like gauge-singlet fermions,
 and denoted that Yukawa coupling of the one field, which corresponds to $S$ in our notation,
 can be always removed rotating vector-like gauge-singlet fermions.}
 and will call $N$ and $S$ right-handed neutrino and sterile neutrino, respectively.

After the EW symmetry breaking,
 the Higgs obtains a nonzero vacuum expectation value, $\langle \Phi \rangle = v/\sqrt{2} \simeq 174\,{\rm GeV}$,
 and then, the Dirac neutrino mass matrix is induced by $m_D = Y_\nu \langle \Phi \rangle$.
In the basis of $n_R=$ ($\nu_L^c$, $N$, $S$)$^T$,
 the neutrino mass terms become
\begin{eqnarray}
	\mathcal{L}_{\rm mass} = -\frac{1}{2} n_R^T C \mathcal{M} n_R + {\rm h.c.}\,,\quad
	{\rm with}\ C=i \gamma^2 \gamma^0\,,
\end{eqnarray} 
 where the $9\times 9$ neutrino mass matrix $\mathcal{M}$ is given by
\begin{eqnarray}
	\mathcal{M} = \left( \begin{array}{ccc}
					0 & m_D^T & 0 \\
					m_D & \mu_N & M_N \\
					0 & M_N^T & \mu_S
				\end{array} \right)\,.
\label{mass_matrix}
\end{eqnarray}
The lepton number is broken by two Majorana masses $\mu_N$ and $\mu_S$,
 and they should be much smaller than $M_N$ and $m_D$
 in the 't Hooft's sense of naturalness.
In the technically natural limit of $\mu_{N,S} \ll m_D, M_N$,
 the order of magnitude of mass eigenvalues are given by
\begin{align}
\label{m_nu}
	\mathcal{O}(m_\nu) &= \frac{m_D^2 \mu_S}{m_D^2+M_N^2} + \mathcal{O}(\mu_{N,S}^2)\,, \\
	\mathcal{O}(m_\pm) &= \pm \sqrt{m_D^2+M_N^2} + \frac{M_N^2 \mu_S}{2(m_D^2+M_N^2)} + \frac{\mu_N}{2} + \mathcal{O}(\mu_{N,S}^2)\,.
\end{align}
In the limit of $\mu_{N,S} \rightarrow 0$,
 active neutrinos become massless and lepton number conservation is restored.
Note that the active neutrino mass $m_\nu$ can be suppressed by small $\mu_S$,
 but does not depend on $\mu_N$ at the tree-level.\footnote{
There are one-loop corrections induced by the $Z$ and Higgs bosons,
 which give a contribution proportional to $\mu_N$,
 and they could be larger than $0.1\,{\rm eV}$~\cite{LopezPavon:2012zg,Dev:2012sg,Dev:2012bd}.
We have checked that our numerical results shown in Sec.~\ref{sec:numerical} avoid such large one-loop corrections.
\label{footnote:one-loop}}
For $m_D \ll M_N$, the heavy neutrino masses are approximated by
 $\mathcal{O}(m_\pm) \simeq \pm M_N + (\mu_S+\mu_N)/2$,
 and thus, the mass difference between them is small as $\mathcal{O}(\mu_{N,S})$.

To explain the active neutrino mass scale $m_\nu \sim 0.1\,{\rm eV}$,
 if there is no accidental cancellation, or no fine-tuning,
 the energy scale of $\mu_S$ should be
\begin{eqnarray}
	\mu_S \sim \left(\frac{M_N}{1 {\rm TeV}}\right)^2 \left(\frac{1}{Y_\nu}\right)^2
		\times 1\,{\rm eV}\,,
\label{mu_scale}
\end{eqnarray}
 for a given $M_N$ and $Y_\nu$.
Technically natural limit of $\mu_{N,S} \ll m_D, M_N$ requires
\begin{eqnarray}
	0.1\,{\rm eV} \lesssim \mu_S \ll 100\,{\rm GeV}\,,
\label{constraint}
\end{eqnarray}
 where the lower and upper bounds have been obtained by
 $m_\nu \sim \mu_S (m_D/M_N)^2 \lesssim \mu_S$,
 and $m_D \lesssim 100\,{\rm GeV}$ corresponding to a perturbativity bound $Y_\nu \lesssim 1$, respectively.
Using Eq.~(\ref{mu_scale}) and $Y_\nu \lesssim 1$,
 $M_N \ll 10^8\,{\rm GeV}$ is required to satisfy Eq.~(\ref{constraint}).
Actually, $\mu_S$ and $M_N$ are constrained more severely by the lepton flavor violation
 as we will show in Sec.~\ref{sec:numerical}.

In the matrix form,
 active neutrino mass matrix is given by
\begin{align}
	m_\nu &\simeq m_D^T (M_N^T)^{-1} \mu_S M_N^{-1} m_D \nonumber \\
		&= m_D^T X^{-1} m_D\,,
\label{active_mass}
\end{align}
 where we have defined $X = M_N \mu_S^{-1} M_N^T$,
 and the last form is the same as the type-I seesaw model~\cite{seesaw}.
The active neutrino mass matrix is diagonalized by Pontecorvo-Maki-Nakagawa-Sakata (PMNS) matrix
 $U_{\rm PMNS}$~\cite{Maki:1962mu,Pontecorvo:1967fh}:
\begin{eqnarray}
	U_{\rm PMNS}^T m_\nu U_{\rm PMNS} = {\rm diag}(m_1, m_2, m_3) \equiv D_m\,,
\end{eqnarray}
 where $m_i$ $(i=1,\, 2,\,{\rm and}\, 3)$ are the mass eigenvalues of three lightest neutrinos.
Actually, since this diagonalization is satisfied at leading order level,
 we will define an unitary matrix diagonalizing the full $9\times9$ neutrino mass matrix in the end of this section,
 and use it for numerical calculations.

Using the Casas-Ibarra parametrization~\cite{Casas:2001sr},
 the neutrino Yukawa coupling matrix is parametrized by
\begin{eqnarray}
	Y_\nu = \frac{\sqrt{2}}{v} V^\dagger \sqrt{D_X} R \sqrt{D_m} U_{\rm PMNS}^\dagger\,,
\label{nuYukawa}
\end{eqnarray}
 where $V$ is a unitary matrix, which diagonalizes $X$ by $V X V^T = {\rm diag}(X_1, X_2, X_3) \equiv D_X$.
The complex orthogonal matrix $R$ can be parameterized by
\begin{eqnarray}
	R = \zeta \left( \begin{array}{ccc}
	c_{12} c_{13} & s_{12} c_{13} & s_{13} \\
	- s_{12} c_{23} - c_{12} s_{23} s_{13} & c_{12} c_{23} - s_{12} s_{23} s_{13} & s_{23} c_{13} \\
	s_{12} s_{23} - c_{12} c_{23} s_{13} & - c_{12} s_{23} - s_{12} c_{23} s_{13} & c_{23} c_{13}
	\end{array} \right)\,,
\end{eqnarray}
 where $c_{ij}=\cos\omega_{ij}$ and $s_{ij}=\sin\omega_{ij}$ with arbitrary complex angles $\omega_{ij}$.
The overall sign $\zeta= \pm \mathbf{1}$ ($\mathbf{1}$ is an $3 \times 3$ identity matrix)
 corresponds to degree of freedom of a parity transformation,
 which determines ${\rm det}[R]=\pm1$ for $\zeta=\pm \mathbf{1}$.
In the following, we assume that $M_N$, $\mu_S$ and $\mu_N$ are diagonal,
 and also $\mu_N=\mu_S$ for simplicity:
 $M_N=D_M \equiv {\rm diag}(M_1,\, M_2,\, M_3)$
 and $\mu_N=\mu_S=D_\mu \equiv {\rm diag}(\mu_1,\, \mu_2,\, \mu_3)$.
Then, $X={\rm diag}(X_1,\, X_2,\, X_3)={\rm diag}(M_1^2/\mu_1,\, M_2^2/\mu_2,\, M_3^2/\mu_3)$,
 and $V$ becomes a unit matrix.

When both $M_N$ and $\mu_{N,S}$ are diagonal,
 before the EW symmetry breaking,
 mass eigenvalues of heavy neutrinos become
\begin{eqnarray}
	m_{i\pm} = \pm M_i + \mu_i \quad {\rm for}\ i=1,\, 2,\, {\rm and}\, 3\,,
\label{heavy}
\end{eqnarray}
 because of $m_D=0$.
% Note that the heavy neutrinos are naturally degenerate above the EW scale.
The mass eigenvectors are given by $N_{i\pm}=(N_i \pm S_i)/\sqrt{2}$,
 and they have the same neutrino Yukawa couplings like
 $(Y_\nu/\sqrt{2}) \overline{N_{i\pm}} \tilde{\Phi}^\dagger \ell$.
Therefore, before the EW symmetry breaking,
 there exist the almost identical particles,
 which have the same couplings and masses with a small difference $\mu_i$.

Since the degenerate heavy neutrinos naturally arise in inverse seesaw models,
 the BAU can be explained by leptogenesis through the neutrino oscillation~\cite{Akhmedov:1998qx,Abada:2015rta}.
On the other hand,
 if the number of sterile neutrinos are larger than the number of right-handed neutrinos,
 there exit additional mass eigenstates with their masses of $\mathcal{O}(\mu_S)$.
The new mass eigenstates can explain
 the short baseline neutrino oscillation anomalies with $\mu_S \sim \mathcal{O}({\rm eV})$~\cite{Kopp:2013vaa},
 and/or the relic density of dark matter with $\mu_S \sim \mathcal{O}({\rm keV})$~\cite{Abada:2014zra}.
However, these issues are beyond the scope of this paper.

For the following discussion,
 it is useful to show the mass matrix in the $(\nu_L^c, N_{i\pm})^T$ basis:
% \begin{eqnarray}
% 	\mathcal{M} = \left( \begin{array}{ccccccc}
% 		0_{3\times 3} & (m_D^{\prime\, T})_{\alpha 1} & (m_D^{\prime\, T})_{\alpha 2} & (m_D^{\prime\, T})_{\alpha 3} & (m_D^{\prime\, T})_{\alpha 1} & (m_D^{\prime\, T})_{\alpha 2} & (m_D^{\prime\, T})_{\alpha 3} \\
% 		(m'_D)_{1 \alpha} & m_{1-} & 0 & 0 & 0 & 0 & 0 \\
% 		(m'_D)_{2 \alpha} & 0 & m_{2-} & 0 & 0 & 0 & 0 \\
% 		(m'_D)_{3 \alpha} & 0 & 0 & m_{3-} & 0 & 0 & 0 \\
% 		(m'_D)_{1 \alpha} & 0 & 0 & 0 & m_{1+} & 0 & 0 \\
% 		(m'_D)_{2 \alpha} & 0 & 0 & 0 & 0 & m_{2+} & 0 \\
% 		(m'_D)_{3 \alpha} & 0 & 0 & 0 & 0 & 0 & m_{3+} \\
% 		\end{array} \right)\,,
% \label{mass_matrix2}
% \end{eqnarray}
\begin{eqnarray}
	\mathcal{M} = \left( \begin{array}{ccccccc}
		0_{3\times 3} & (m_D^{\prime\, T})_{\alpha 1} & (m_D^{\prime\, T})_{\alpha 1} & (m_D^{\prime\, T})_{\alpha 2} & (m_D^{\prime\, T})_{\alpha 2} & (m_D^{\prime\, T})_{\alpha 3} & (m_D^{\prime\, T})_{\alpha 3} \\
		(m'_D)_{1 \alpha} & m_{1-} & 0 & 0 & 0 & 0 & 0 \\
		(m'_D)_{1 \alpha} & 0 & m_{1+} & 0 & 0 & 0 & 0 \\
		(m'_D)_{2 \alpha} & 0 & 0 & m_{2-} & 0 & 0 & 0 \\
		(m'_D)_{2 \alpha} & 0 & 0 & 0 & m_{2+} & 0 & 0 \\
		(m'_D)_{3 \alpha} & 0 & 0 & 0 & 0 & m_{3-} & 0 \\
		(m'_D)_{3 \alpha} & 0 & 0 & 0 & 0 & 0 & m_{3+} \\
		\end{array} \right)\,,
\label{mass_matrix2}
\end{eqnarray}
 where $m'_D = m_D/\sqrt{2}$.
Now, we define $U$ as an unitary matrix diagonalizing this full $9 \times 9$ mass matrix,
 which is given in the following form:
\begin{eqnarray}
	U^T \mathcal{M} U = \mathcal{M}_{\rm diag}\quad {\rm with}\quad 
	\left( \nu_e,\, \nu_\mu,\, \nu_\tau,\, N_{1-},\, N_{1+},\, N_{2-},\, N_{2+},\, N_{3-},\, N_{3+} \right)^T
	= U_{\alpha i}\, \nu_i \,.
\end{eqnarray}
The mass eigenstates $\nu_1$, $\nu_2$ and $\nu_3$ correspond to the active neutrinos,
 and $\nu_i$'s ($i=4\sim9$) are the heavy neutrinos in light order from $i=4$ to 9. 
For $m_D \ll M_N$,
 the heavy mass eigenstates are almost composed of $N_{i\pm}$.
In particular, diagonal elements $U_{ii}$ ($i=4\sim9$) are nearly unity,
 unless $M_N$ is degenerate.
In addition,
 matrix elements expressing the left-right mixing are almost satisfy
 $U_{\alpha 4}^2 \simeq U_{\alpha 5}^2$,  $U_{\alpha 6}^2 \simeq U_{\alpha 7}^2$ and $U_{\alpha 8}^2 \simeq U_{\alpha 9}^2$
 for $\alpha = e,\, \mu\, {\rm and}\, \tau$,
 which are exactly satisfied in the limit of $\mu_i \to 0$.

%%%%%%%%%%%%%%%%%%%%%%%%%%%%%%%%%%%%%%
\section{Neutrinoless double beta decay} \label{sec:beta}
%%%%%%%%%%%%%%%%%%%%%%%%%%%%%%%%%%%%%%

The massive Majorana neutrinos induce neutrinoless double beta decay,
 and its rate is proportional to the squared of the effective neutrino mass,
 which is given by
\begin{eqnarray}
	m_{\rm eff} = \left| \sum_{i=1}^9 U_{ei}^2 \frac{\bar{p}^2}{\bar{p}^2 + m_i^2} m_i \right| 
		\simeq \left| \left( \sum_{i=1}^3 U_{ei}^2 m_i  \right) + m_{\rm eff}^N \right| \,,
\label{meff}
\end{eqnarray}
 where $\bar{p}^2 \sim (200\,{\rm MeV})^2$ is the typical virtual momentum of the neutrino.
The first summation term corresponds to contributions from the active neutrinos,
 and $m_{\rm eff}^N$ stands for the heavy neutrino contributions.
In the inverse seesaw model,
 $m_{\rm eff}^N$ can be obtained by~\cite{Abada:2014vea,Abada:2014nwa}
\begin{align}
	m_{\rm eff}^N \simeq& \sum_{i=1}^3 \left[
		- U_{e(2i+2)}^2 \frac{\bar{p}^2}{\bar{p}^2 + m_{i-}^2} |m_{i-}|
		+ U_{e(2i+3)}^2 \frac{\bar{p}^2}{\bar{p}^2 + m_{i+}^2} |m_{i+}|
	\right] \nonumber\\
	=& - \sum_{i=1}^3 \left[ \left( U_{e(2i+2)}^2 + U_{e(2i+3)}^2 \right)
		\frac{[(M_i^2 - \mu_i^2) - \bar{p}^2] \bar{p}^2}{[(M_i^2 - \mu_i^2) + \bar{p}^2]^2 + 4 \bar{p}^2 \mu_i^2}\, \mu_i \right. \nonumber\\
	&\left. \quad \quad + \left( U_{e(2i+2)}^2 - U_{e(2i+3)}^2 \right)
		\frac{[(M_i^2 - \mu_i^2) + \bar{p}^2] \bar{p}^2}{[(M_i^2 - \mu_i^2) + \bar{p}^2]^2 + 4 \bar{p}^2 \mu_i^2}\, M_i \right] \,.
\label{meff_N}
\end{align}
In the limit of $\mu_i \to 0$ (no lepton number asymmetry),
 the heavy neutrino contributions exactly vanish due to 
 $U_{e 4}^2 = U_{e 5}^2$,  $U_{e 6}^2 = U_{e 7}^2$ and $U_{e 8}^2 = U_{e 9}^2$.
Notice that we can usually take $(M_i^2-\mu_i^2)^{1/2} \simeq M_i$,
 but this approximation becomes invalid
 if one considers leptogenesis thorough neutrino oscillations to explain the BAU,
 in which $0.1 \lesssim 2\mu_i/M_i \lesssim 1$~\cite{Abada:2015rta}.

\begin{figure}[t]
\centering
	\includegraphics[scale=0.83,clip]{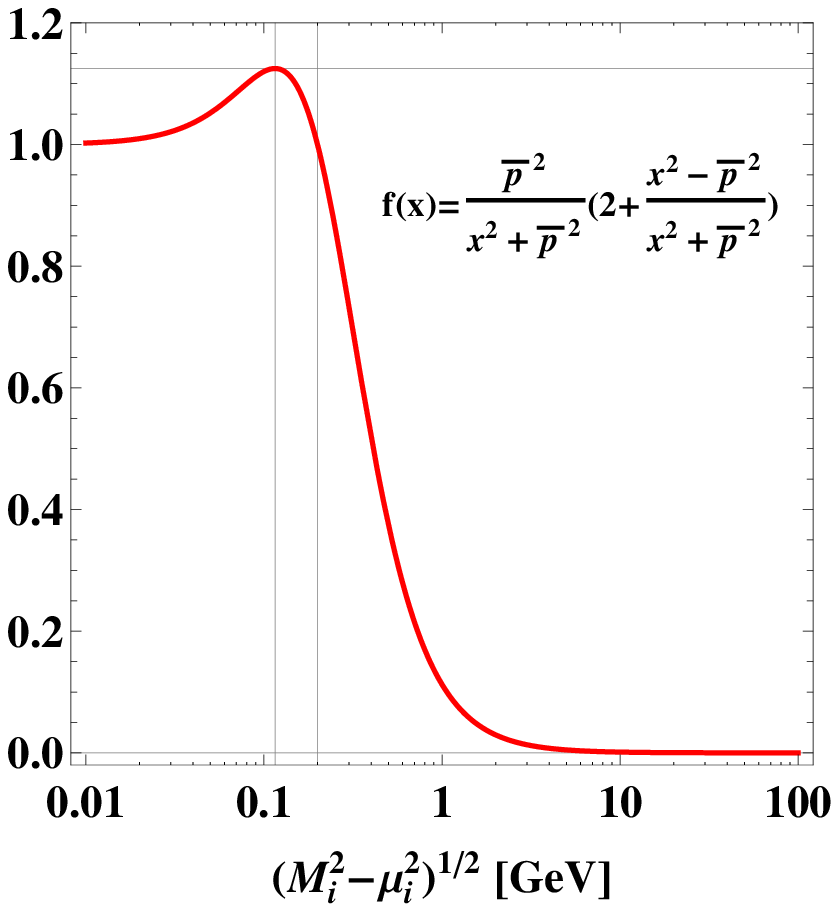} \hspace{2mm}
	\includegraphics[scale=0.83,clip]{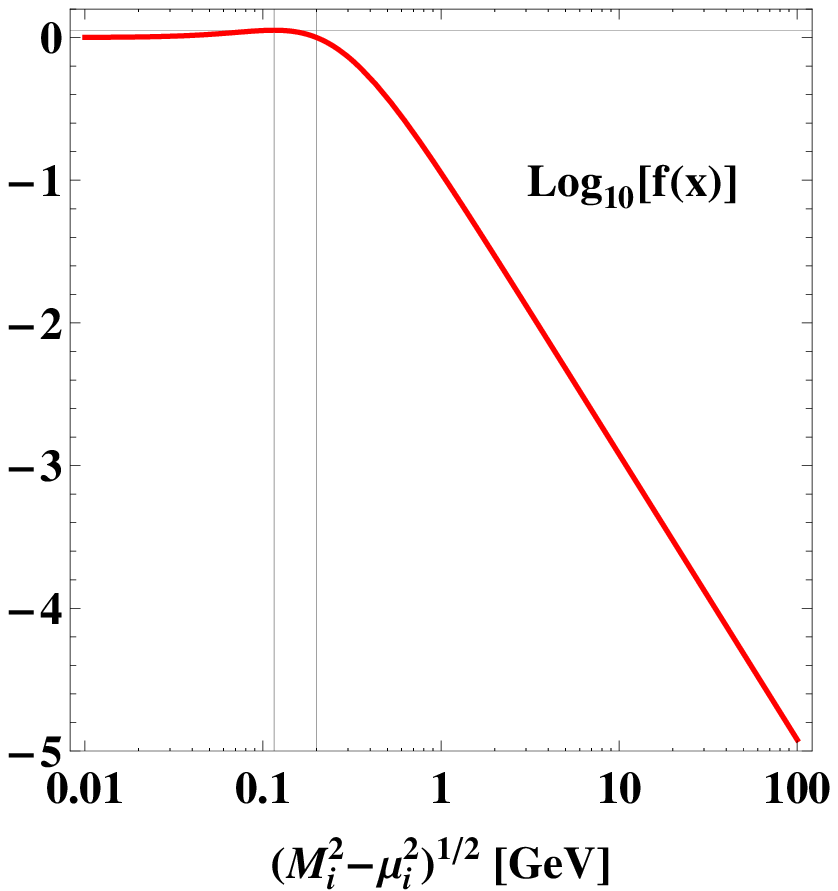}
\caption{The suppression factor in Eq.~(\ref{meff_N_fin}).
The vertical lines show $\sqrt{\bar{p}^2}=200\,{\rm MeV}$ and $\sqrt{\bar{p}^2/3}$.
The horizontal line show $9/8$ as the maximal value.
The right panel is shown with a logarithmic scale.
}
\label{function}
\end{figure}

% Using the unitarity relation $\sum_{i=1}^9 U_{\alpha i} U_{\beta i} m_i = 0$
%  ($\alpha,\, \beta = e,\, \mu,\, {\rm and}\, \tau$), which induces
% \begin{eqnarray}
% 	\left( \sum_{i=1}^3 U_{\alpha i}^2 m_i \right)
% 	+ \sum_{i=1}^3 \left[ \left( U_{\alpha (2i+2)}^2 + U_{\alpha (2i+3)}^2 \right) \mu_i
% 	- \left( U_{\alpha (2i+2)}^2 - U_{\alpha (2i+3)}^2 \right) M_i \right] = 0\,,
% \end{eqnarray}
%  Eq.~(\ref{meff_N}) can be rewritten by
% \begin{align}
% 	m_{\rm eff}^N 
% 	\simeq& \sum_{i=1}^3 \left[ 
% 		\frac{[(M_i^2 - \mu_i^2) - \bar{p}^2] \bar{p}^2}{[(M_i^2 - \mu_i^2) + \bar{p}^2]^2}\, U_{\alpha i}^2 m_i
% 	- \left( U_{e(2i+2)}^2 - U_{e(2i+3)}^2 \right)
% 		\frac{2 (M_i^2 - \mu_i^2) \bar{p}^2}{[(M_i^2 - \mu_i^2) + \bar{p}^2]^2}\, M_i \right] \,,
% \end{align}
%  where we have neglected $\bar{p}^2 \mu_i^2$ term in the denominator.
In addition, the unitary matrix $U$ has the following approximate relations: 
\begin{eqnarray}
	U_{\alpha (2i+2)}^2 + U_{\alpha (2i+3)}^2 \simeq \frac{(m_D^*)_{i \alpha}^2}{M_i^2}\quad {\rm and}\quad
	U_{\alpha (2i+2)}^2 - U_{\alpha (2i+3)}^2 \simeq \frac{(m_D^*)_{i \alpha}^2}{M_i^2} \frac{2 \mu_i}{M_i}\,,
\end{eqnarray}
 where $\alpha = e,\, \mu\, {\rm and}\, \tau$.
Then, we find the analytic form of $m_{\rm eff}^N$ as
\begin{align}
	m_{\rm eff}^N \simeq - \sum_{i=1}^3  
	\frac{\bar{p}^2}{(M_i^2 - \mu_i^2) + \bar{p}^2}
	\left[ 2 + \frac{(M_i^2 - \mu_i^2) - \bar{p}^2}{(M_i^2 - \mu_i^2) + \bar{p}^2} \right] (\tilde{m}_\nu^*)_{ie} \,,
\label{meff_N_fin}
\end{align}
with
\begin{align}
	(\tilde{m}_\nu)_{ie} = \frac{(m_D)_{ie}^2 \mu_i}{M_i^2}\,.
\end{align}
% \begin{align}
% 	m_{\rm eff}^N 
% 	\simeq& \sum_{i=1}^3 \left[ 
% 		\frac{[(M_i^2 - \mu_i^2) - \bar{p}^2] \bar{p}^2}{[(M_i^2 - \mu_i^2) + \bar{p}^2]^2}\, U_{\alpha i}^2 m_i
% 	- \frac{4 (M_i^2 - \mu_i^2) \bar{p}^2}{[(M_i^2 - \mu_i^2) + \bar{p}^2]^2}\, (\tilde{m}_\nu^*)_{ie} \right]\,,
% \end{align}
% The first term is lower than the active neutrino contribution.
The coefficient factor in Eq.~(\ref{meff_N_fin}) is strongly suppressed for a large $(M_i^2-\mu_i^2)^{1/2}$.
Figure~\ref{function} shows it as a function of $(M_i^2-\mu_i^2)^{1/2}$.
From Fig.~\ref{function},
 we can see that sizable $m_{\rm eff}^N$ is likely to be obtained by $(M_i^2-\mu_i^2)^{1/2} \lesssim 1\,{\rm GeV}$.
Since $(\tilde{m}_\nu)_{ie}$ appears in the active neutrino mass matrix,
 i.e. $(m_\nu)_{ee} \simeq \sum_{i=1}^3 (\tilde{m}_\nu)_{ie}$ [see Eq.~(\ref{active_mass})],
 $(\tilde{m}_\nu)_{ie}$ is typically $\mathcal{O}(m_\nu)$ without any fine-tuning.
However, it can be much larger than $\mathcal{O}(m_\nu)$ with a sufficient fine-tuning,
 so that $m_{\rm eff}^N$ can be much larger than the naive estimation.
We will show numerical results in Sec.~\ref{sec:numerical}.

The existence of heavy neutrinos also induce non-standard interactions in the leptonic sector,
 which correspond to non-unitarity of the PMNS matrix.
The deviation from unitarity can be estimated by
\begin{eqnarray}
	\epsilon_{\alpha \beta} \equiv \left| \sum_{i=4}^{7} U_{\alpha i} U^*_{\beta i}\right|
		= \left| \delta_{\alpha \beta} - \left( N N^\dagger \right)_{\alpha \beta} \right|\,,
\end{eqnarray}
 where $N$ is the $3 \times 3$ non-unitary matrix
 describing the mixing between the light neutrino mass eigenstates and the $SU(2)_L$ gauge eigenstates,
 that is, the PMNS matrix.
The values of $\epsilon_{\alpha \beta}$ are severely constrained by the combined data from
 neutrino oscillation data,
 lepton-flavor-violating decays of charged leptons,
 non-universality of weak interaction,
 CKM unitarity bounds,
 and EW precision data~\cite{Fernandez-Martinez:2016lgt}:
\begin{eqnarray}
	|\epsilon_{\alpha \beta}| \leq \left( \begin{array}{ccc}
		2.5 \times 10^{-3} & 2.4 \times 10^{-5} & 2.7 \times 10^{-3} \\
		2.4 \times 10^{-5} & 4.0 \times 10^{-4} & 1.2 \times 10^{-3} \\
		2.7 \times 10^{-3} & 1.2 \times 10^{-3} & 5.6 \times 10^{-3} 
	\end{array} \right)\,.
\label{non-unitarity}
\end{eqnarray}
The constraint on $\epsilon_{e\mu}$ ($=\epsilon_{\mu e}$),
 which comes from a constraint on the lepton-flavor-violating muon decay $\mu \to e \gamma$,
 is much stronger than the others.
Without assuming accidental cancellation or special textures for the neutrino Yukawa coupling matrix
 (that is, all components of Yukawa coupling matrix have the same order values),
 once model parameters are set to satisfy $\epsilon_{e\mu}\leq 2.4 \times 10^{-5}$,
 the other constraints can be simultaneously satisfied.

% \newpage
%%%%%%%%%%%%%%%%%%%%%%%%%%%%%%%%%%%%%%
\section{Higgs mass correction} \label{sec:mass_corr}
%%%%%%%%%%%%%%%%%%%%%%%%%%%%%%%%%%%%%%

The heavy neutrinos may lead a sizable Higgs mass correction.
In our notation, the Higgs potential is given by $V=\lambda (\Phi^\dagger \Phi)^2 + m_H^2 \Phi^\dagger \Phi$,
 and then, the Higgs mass is obtained by $M_h^2=-2m_H^2=(125\,{\rm GeV})^2$.
Then, the heavy neutrinos contribute the Higgs mass parameter as
\begin{align}
	\delta m_H^2 &= -\frac{2}{16 \pi^2}\, {\rm Tr}[Y_\nu^\dagger M_N^T M_N Y_\nu] \ln \left(\frac{M_{\rm Pl}^2}{M_N^2} \right) \nonumber \\
	&= -\frac{4}{16 \pi^2 v^2}\, {\rm Tr}[D_M^2 D_X R D_m R^\dagger] \ln \left(\frac{M_{\rm Pl}^2}{M_N^2} \right)\,,
\label{mass_corr}
\end{align}
 where we have used Casas-Ibarra parametrization~(\ref{nuYukawa}),
 and assumed both $M_N$ and $\mu_S$ are diagonal.
We have taken cutoff scale as the reduced Planck scale $M_{\rm Pl}=2.44\times 10^{18}\,{\rm GeV}$.
In the last expression of Eq.~(\ref{mass_corr}),
 $D_X \to D_M$ corresponds to the Higgs mass correction in the type-I seesaw model.
Thus, in the inverse seesaw model,
 the Higgs mass correction can be enhanced by a factor $M_N/\mu_S$ compared to the type-I seesaw model.

Since we have assumed that both $M_N$ and $\mu_S$ are diagonal,
 the trace part in Eq.~(\ref{mass_corr}) can be simply shown by
\begin{align}
	{\rm Tr}[D_M^2 D_X R D_m R^\dagger] = 
	\left( \frac{M_1^4}{\mu_1},\, \frac{M_2^4}{\mu_2},\, \frac{M_3^4}{\mu_3} \right)
	\left( \begin{array}{ccc}
	|R_{11}|^2 & |R_{12}|^2 & |R_{13}|^2 \\
	|R_{21}|^2 & |R_{22}|^2 & |R_{23}|^2 \\
	|R_{31}|^2 & |R_{32}|^2 & |R_{33}|^2 
	\end{array} \right)
	\left( \begin{array}{c}
	m_1 \\ m_2 \\ m_3
	\end{array} \right)\,.
\label{trace}
\end{align}
In addition, when both $M_N$ and $\mu_S$ are degenerate,
 i.e., $M_N=M_d \mathbf{1}$ and $\mu_S=\mu_d \mathbf{1}$,
 this trace can be rewritten by
\begin{align}
	{\rm Tr}[D_M^2 D_X R D_m R^\dagger] = \frac{M_d^4}{\mu_d} \left( R_1 m_1 + R_2 m_2 + R_3 m_3 \right)\,,
\end{align}
with $R_i \equiv \sum_{j=1}^3 |R_{ji}|^2$, which is obtained by
\begin{align}
	R_1 &= |c_{12}|^2 |c_{13}|^2 + (|s_{12}|^2+|c_{12}|^2 |s_{13}|^2) \cosh (2 {\rm Im}[\omega_{23}])
			+ 2 {\rm Im}[s_{12} c_{12}^* s_{13}^*] \sinh (2 {\rm Im}[\omega_{23}])\,, \nonumber\\
	R_2 &= |s_{12}|^2 |c_{13}|^2 + (|c_{12}|^2+|s_{12}|^2 |s_{13}|^2) \cosh (2 {\rm Im}[\omega_{23}])
			+ 2 {\rm Im}[s_{12} c_{12}^* s_{13}] \sinh (2 {\rm Im}[\omega_{23}])\,, \nonumber\\
	R_3 &= |s_{13}|^2 + |c_{13}|^2 \cosh (2 {\rm Im}[\omega_{23}])\,.
\end{align}
In this expression, we can see $R_i \geq 1$.
This fact is easily understood in two flavor case,
 in which the number of right-handed neutrinos and sterile neutrinos are two.
In the two flavor case,
 the lightest neutrino is massless,
 and $R$ is expressed by only one complex angle.
The normal hierarchy (NH) corresponds to
 $m_1=0$ and $\omega_{12}=\omega_{13}=0$,
 which lead $R_2 = R_3 = \cosh (2 {\rm Im}[\omega_{23}]) \geq 1$.
In the same way,
 the inverted hierarchy (IH) corresponds to
 $m_3=0$ and $\omega_{13}=\omega_{23}=0$,
 which lead $R_1 = R_2 = |c_{12}|^2+|s_{12}|^2 = \cosh (2 {\rm Im}[\omega_{12}]) \geq 1$.

\begin{figure}[t]
\centering
	\includegraphics[scale=1,clip]{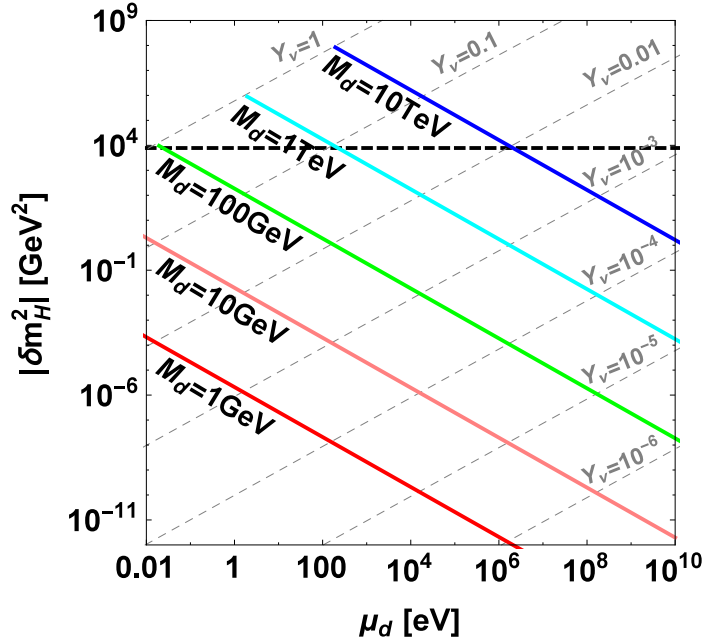}
\caption{Higgs mass correction as a function of $\mu_d$,
 which are shown by the red, pink, green, cyan and blue lines for
 $M_d=1,\, 10,\, 10^2,\, \cdots,\, {\rm and}\, 10^5\,{\rm GeV}$, respectively.
The black-dashed line corresponds to $|\delta m_H^2| =M_h^2/2$ with $M_h=125\,{\rm GeV}$.
Typical values of neutrino Yukawa coupling can be estimated by Eq.~(\ref{mu_scale}),
 and they are shown by the gray-dashed lines for
 $Y_\nu=1,\, 10^{-1},\, 10^{-2},\, \cdots,\, {\rm and}\, 10^{-6}$ from the upper left to the lower right.
}
\label{mu_dep}
\end{figure}

As a result, the Higgs mass correction has the minimal value:
\begin{align}
	|\delta m_H^2| \geq \frac{4 M_d^4}{16 \pi^2 v^2 \mu_d} \left( m_1 + m_2 + m_3 \right) \ln \left(\frac{M_{\rm Pl}^2}{M_d^2} \right) \,,
\label{mass_corr2}
\end{align}
 where the equals sign holds with, e.g., $R=\mathbf{1}$.
For the non-degenerate case,
 $M_d$ is replaced by the heaviest neutrino mass.
Figure\,\ref{mu_dep} shows the Higgs mass correction as a function of $\mu_d$,
 which corresponds to the minimal value of Eq.\,(\ref{mass_corr2}).
For reference values,
 we have considered the NH case,
 and taken $m_1=0$, $m_2= \sqrt{7.49 \times 10^{-5}}\,{\rm eV}$
 and $m_3= \sqrt{2.484 \times 10^{-3}}\,{\rm eV}$~\cite{Gonzalez-Garcia:2014bfa}.
The red, pink, green, cyan and blue lines show $|\delta m_H^2|$ for
 $M_d=1,\, 10,\, 10^2,\, \cdots,\, {\rm and}\, 10^5\,{\rm GeV}$, respectively.
Using the seesaw relation~(\ref{mu_scale}),
 typical values of neutrino Yukawa coupling can be determined by a function of $\mu_d$ with a fixed $M_d$,
 which are shown in the gray-dashed lines. 
From Fig.~\ref{mu_dep},
 we can see that the Higgs mass correction becomes larger as $\mu_d$ becomes smaller.
When $\mu_d$ is smaller than $0.1\,{\rm eV}$,
 the inverse seesaw mechanism does not work successfully as shown in Eq.~(\ref{constraint}).
For $\mu_d > 0.1\,{\rm eV}$,
 the heavy neutrino contributions to the Higgs mass become dominant for $M_d \gtrsim 160\,{\rm GeV}$.
This fact is much different from the type-I seesaw model,
 in which $|\delta m_H^2| < M_h^2/2$ for $M_d < 10^6\,{\rm GeV}$.
Therefore, from the naturalness point of view,
 inverse seesaw models have to introduce heavy neutrinos more carefully than the type-I seesaw model.

% \newpage
%%%%%%%%%%%%%%%%%%%%%%%%%%%%%%%%%%%%%%
\section{Numerical analysis} \label{sec:numerical}
%%%%%%%%%%%%%%%%%%%%%%%%%%%%%%%%%%%%%%

In this section,
 we show numerical results of the effective neutrino mass (\ref{meff}) and the Higgs mass correction (\ref{mass_corr}).
We focus on the case where $M_N$ and $\mu_N=\mu_S$ are real diagonal matrices and normally hierarchical $M_N$, i.e., $M_1 < M_2 < M_3$.
Actually, even if they are not normally hierarchical,
 we have found similar conclusions for other hierarchies.
In our numerical calculations,
 we take the input parameters as
\begin{align}
	m_{1\,(3)} &= [10^{-4}\,{\rm eV},\ 0.07\,(0.065)\,{\rm eV}]\ {\rm for\ NH\ (IH)}\,,\nonumber\\
	M_i &= [10\,{\rm MeV},\ 100\,{\rm TeV}]\,,\nonumber\\
	\mu_i &= [1\,{\rm eV},\ 1\,{\rm MeV}]\,,\nonumber\\
	\omega_{ij} &= [0,\ \pi] \times e^{i\,[0,\ 2\pi]}\,,
\end{align}
 and neutrino oscillation parameters satisfy the current experimental constraints shown in Table~\ref{nu_data}.
\begin{table}[t]
\centering
\begin{tabular}{|c|ccccc|}\hline
& $\sin^2 \theta_{12}$ & $\sin^2 \theta_{23}$ & $\sin^2 \theta_{13}$ & $\frac{\Delta m_{21}^2}{10^{-5}{\rm eV}^2}$ & $\frac{\Delta m_{3\ell}^2}{10^{-3}{\rm eV}^2}$ \\
\hline
NH & $0.273 \to 0.349$ & $0.390 \to 0.639$ & $0.0187 \to 0.0250$ & $7.02 \to 8.08$ & $2.351 \to 2.618$ \\
IH & $0.273 \to 0.349$ & $0.400 \to 0.637$ & $0.0190 \to 0.0251$ & $7.02 \to 8.08$ & $2.341 \to 2.595$ \\
\hline
\end{tabular}
\caption{Global fit values of neutrino oscillation parameters in 3$\sigma$ CL range~\cite{Gonzalez-Garcia:2014bfa}.
The mass squared differences are defined by
 $\Delta m_{21}^2 = m_2^2 - m_1^2$ and $\Delta m_{3\ell}^2 = |m_3^2 - m_\ell^2|$
 with $\ell = 1$ and 2 for the NH and the IH cases, respectively.
There are no constraints for all CP phases in 3$\sigma$ CL range.}
\label{nu_data}
\end{table}
The upper bound of the lightest active neutrino mass is given by the cosmological bound
 $\sum m_i < 0.23\,{\rm eV}$~\cite{Ade:2015xua}.
In addition, the mixing matrix $U$ satisfies the constraints of the non-unitarity (\ref{non-unitarity}).
Note that, however, there exist severe constraints in a low $M_i$ region ($M_i \lesssim 2\,{\rm GeV}$),
 which are obtained by $\pi$ and $K$ peak searches, $\pi$, $K$, $D$, $Z$ decay searches,
 and LHC collider searches.
They are summarized in Ref.~\cite{Atre:2009rg},
 and we also apply their constraints in our analysis.

% \begin{eqnarray}
% 	|U_{\alpha i}|_{3\sigma} = \left( \begin{array}{ccc}
% 		0.798 \to 0.843 & 0.517 \to 0.584 & 0.137 \to 0.158 \\
% 		0.231 \to 0.518 & 0.441 \to 0.693 & 0.617 \to 0.790 \\
% 		0.251 \to 0.530 & 0.468 \to 0.711 & 0.595 \to 0.773 
% 	\end{array} \right)\,.
% \end{eqnarray}

\begin{figure}[t]
\centering
	\includegraphics[scale=0.83,clip]{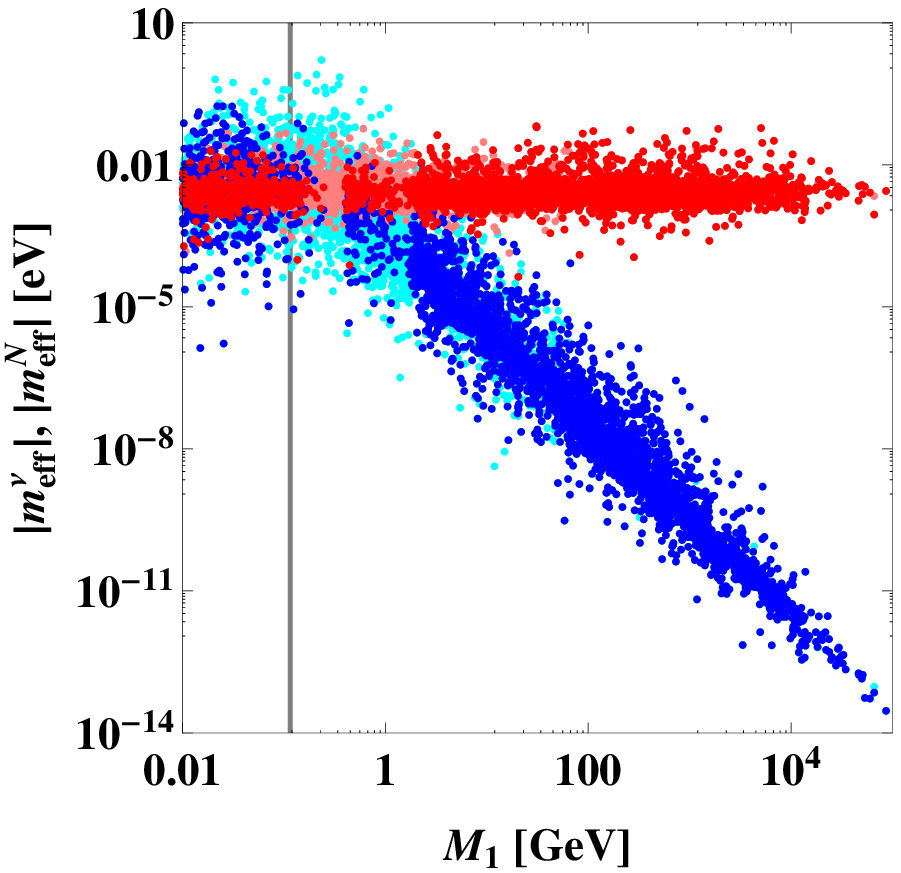} \hspace{2mm}
	\includegraphics[scale=0.83,clip]{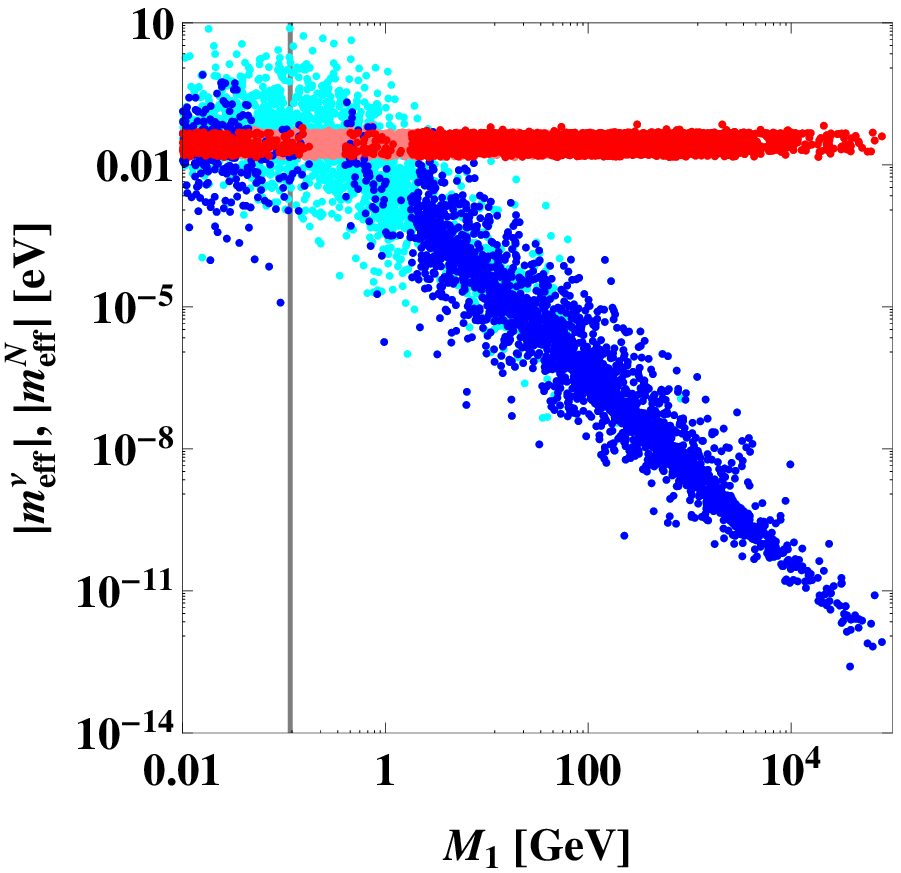}
\caption{Heavy neutrino mass dependence of $|m_{\rm eff}^N|$ (blue dots).
The red dots show $|m_{\rm eff}^\nu|$, which does not depend on heavy neutrinos.
The cyan and pink dots show the excluded points due to the constraints in Ref.~\cite{Atre:2009rg}
 for $|m_{\rm eff}^N|$ and $|m_{\rm eff}^\nu|$, respectively.
The vertical line shows $\sqrt{\bar{p}^2/3}$ with $\sqrt{\bar{p}^2}=200\,{\rm MeV}$.
The left and right panels correspond to the NH and the IH cases, respectively.
}
\label{fig:meffN}
\end{figure}

For the neutrinoless double beta decay,
 the lightest heavy neutrino mass $M_1$ is sensitive.
Figure~\ref{fig:meffN} shows $M_1$ dependence of $|m_{\rm eff}^N|$
 with the blue dots,
 while the red dots show the active neutrino contribution $|m_{\rm eff}^\nu|$,
 which does not depend on heavy neutrinos.
The cyan and pink dots show the excluded points due to the constraints in Ref.~\cite{Atre:2009rg}
 for $|m_{\rm eff}^N|$ and $|m_{\rm eff}^\nu|$, respectively.
The behavior of $|m_{\rm eff}^N|$ is the same as Fig.~\ref{function},
 which is expected from our analytical result (\ref{meff_N_fin}).
The heavy neutrino contribution can be much larger than the active neutrino contribution
 in the range of $M_1 \lesssim 1\,{\rm GeV}$.
\begin{figure}[!ht]
\centering
	\includegraphics[scale=0.83,clip]{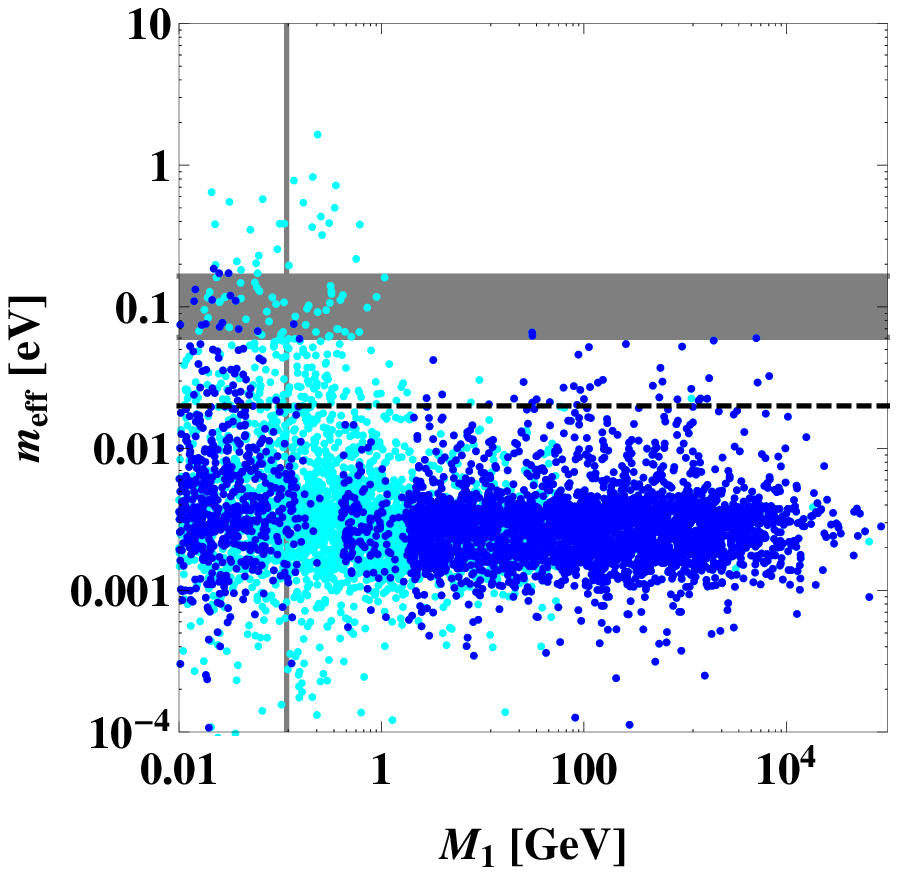} \hspace{2mm}
	\includegraphics[scale=0.83,clip]{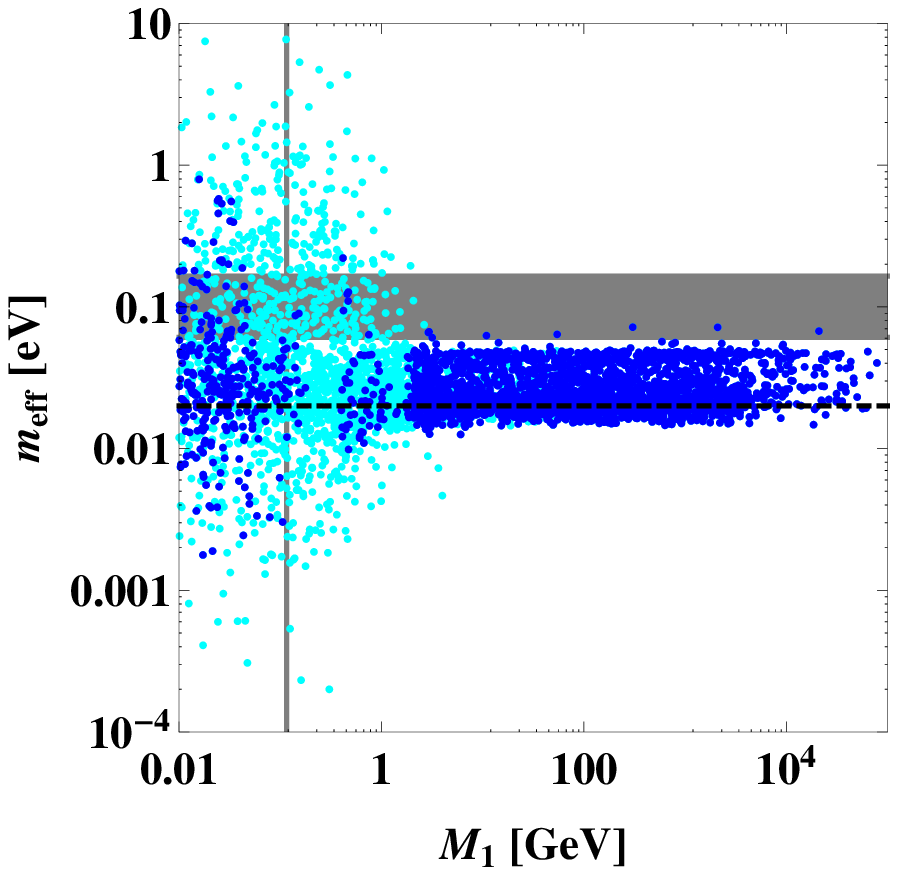}
\caption{Heavy neutrino mass dependence of $m_{\rm eff}$ (blue dots).
The cyan dots show the excluded points due to the constraints in Ref.~\cite{Atre:2009rg}.
The gray band and the black-dashed line show
 the current upper bound $m_{\rm eff} < (61-165)\,{\rm meV}$
 and the future sensitivity $m_{\rm eff} < 0.02\,{\rm eV}$, respectively.
The vertical line shows $\sqrt{\bar{p}^2/3}$ with $\sqrt{\bar{p}^2}=200\,{\rm MeV}$.
The left and right panels correspond to the NH and the IH cases, respectively.
}
\label{fig:meff}
\end{figure}

Figure~\ref{fig:meff} shows $M_1$ dependence of $m_{\rm eff}$ (blue dots).
The cyan dots show the excluded points due to the constraints in Ref.~\cite{Atre:2009rg}.
The gray band and the black-dashed line corresponds to
 the current upper bound $m_{\rm eff} < (61-165)\,{\rm meV}$ obtained by the KamLAND-Zen experiment~\cite{KamLAND-Zen:2016pfg}
 and their future sensitivity $m_{\rm eff} < 0.02\,{\rm eV}$, respectively.
For $M_1 > 1\,{\rm GeV}$,
 the heavy neutrino contribution is strongly suppressed,
 and thus, there are almost no points above the current upper bound.
Note that there exists $m_{\rm eff}<|m_{\rm eff}^\nu|$ region,
 since $m_{\rm eff}^\nu$ and $m_{\rm eff}^N$ can be canceled each other.
Thus, the IH case is not completely excluded.

\begin{figure}[t]
\centering
	\includegraphics[scale=0.83,clip]{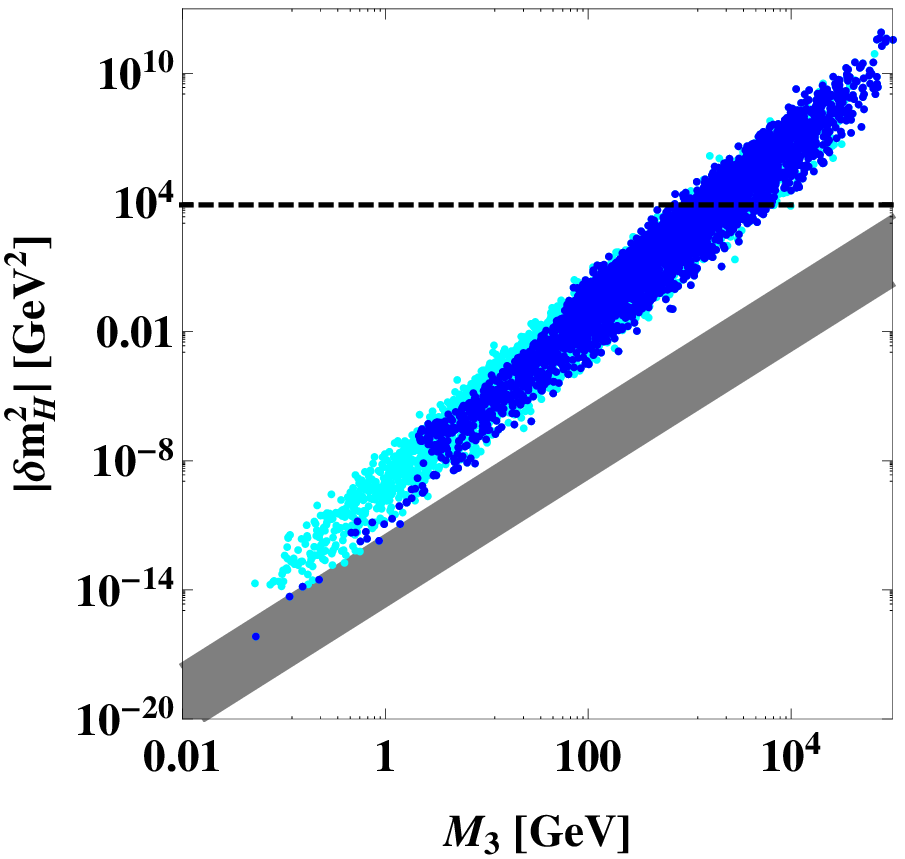} \hspace{2mm}
	\includegraphics[scale=0.83,clip]{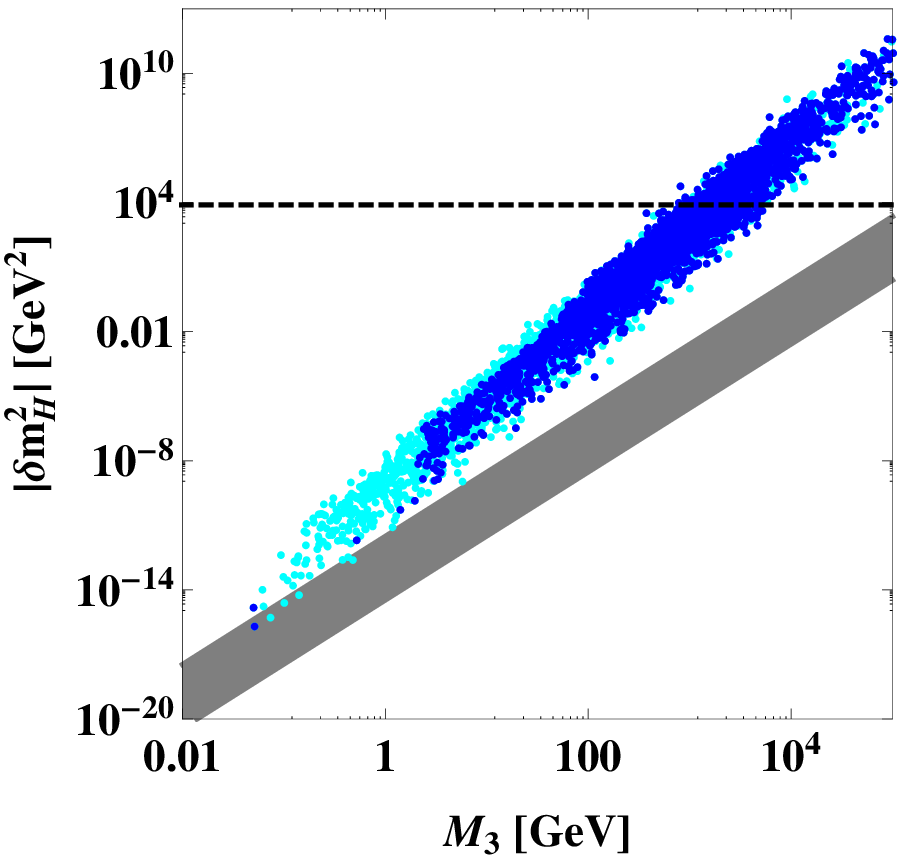}
\caption{Heavy neutrino mass dependence of the Higgs mass correction (blue dots).
The cyan dots show the excluded points due to the constraints in Ref.~\cite{Atre:2009rg}.
The gray band and the black-dashed line show the type-I seesaw case,
 and $|\delta m_H^2| =M_h^2/2$ with $M_h=125\,{\rm GeV}$, respectively.
The left and right panels correspond to the NH and the IH cases, respectively.
}
\label{fig:mH}
\end{figure}

\begin{figure}[t]
\centering
	\includegraphics[scale=0.83,clip]{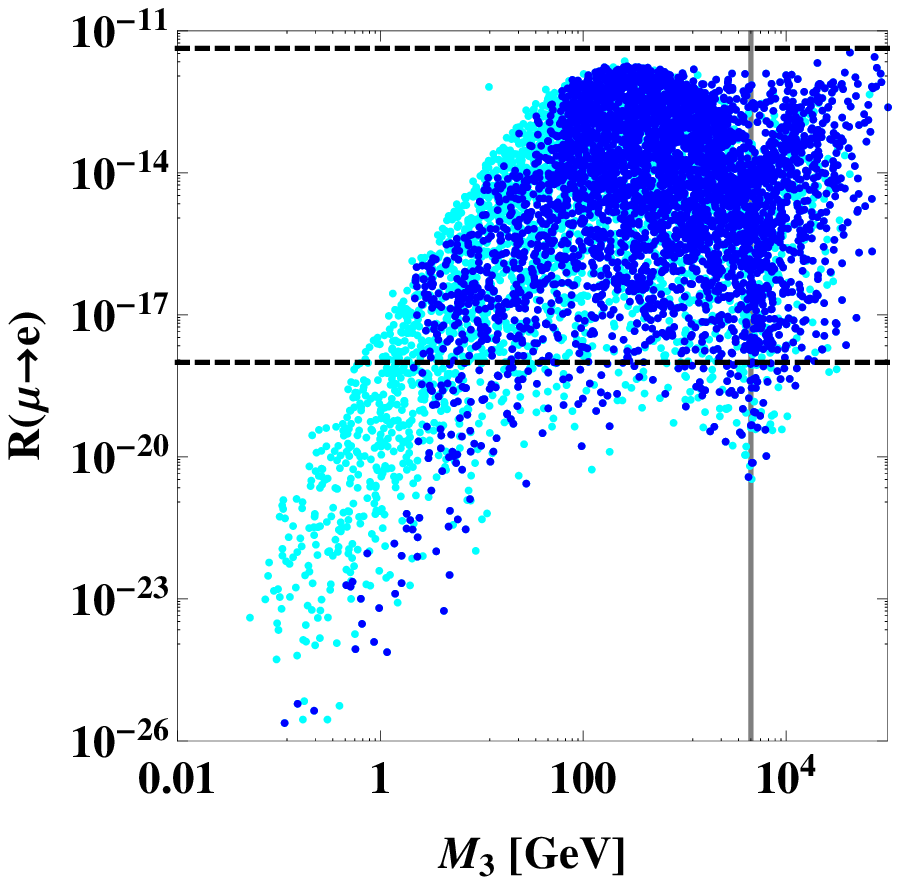} \hspace{2mm}
	\includegraphics[scale=0.83,clip]{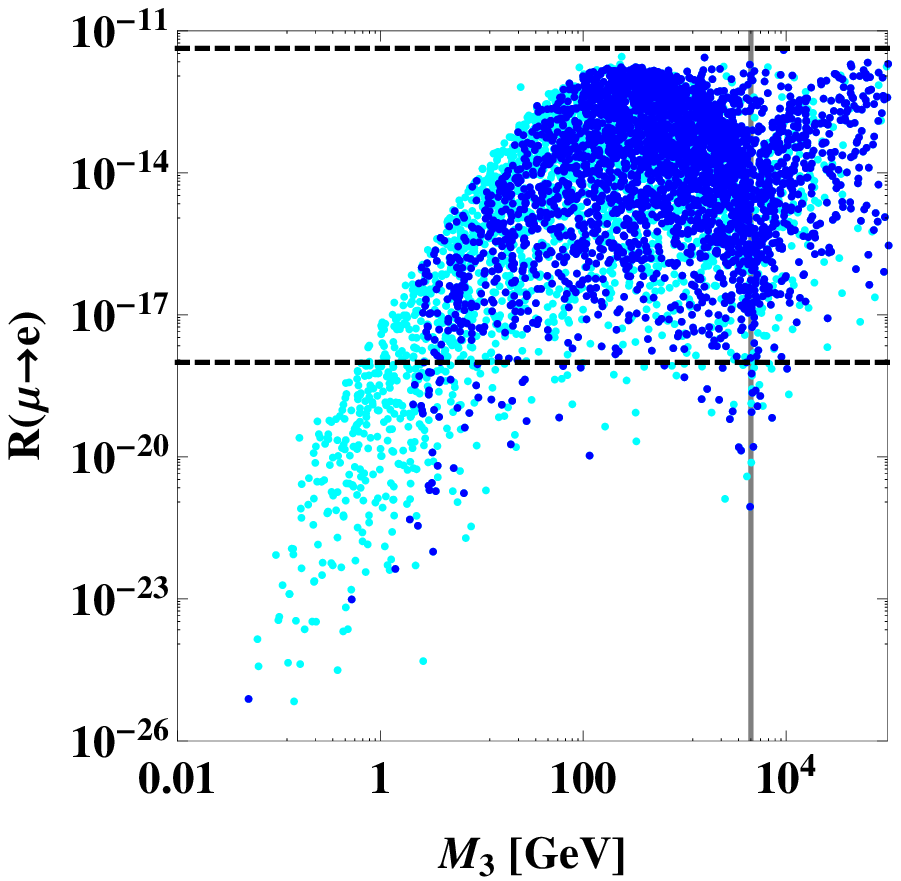}
\caption{The rate of $\mu \to e$ conversion in Titanium (blue dots).
The cyan dots show the excluded points due to the constraints in Ref.~\cite{Atre:2009rg}.
The black-dashed lines correspond to the current upper bound $R_{\mu \to e} < 4.3\times 10^{-12}$
 and the future sensitivity of PRISM experiment $R_{\mu \to e} < 10^{-18}$.
The vertical line shows $M_3=4.5\,{\rm TeV}$,
 at which the contribution of the $\mu \to e$ conversion vanishes.
The left and right panels correspond to the NH and the IH cases, respectively.
}
\label{fig:mue}
\end{figure}

For the Higgs mass correction,
 the heaviest heavy neutrino mass $M_3$ is sensitive.
Figure~\ref{fig:mH} shows $M_3$ dependence of $|\delta m_H^2|$ (blue dots).
The cyan dots show the excluded points due to the constraints in Ref.~\cite{Atre:2009rg}.
The gray band shows the type-I seesaw case,
 and the black-dashed line corresponds to $|\delta m_H^2| =M_h^2/2$ with $M_h=125\,{\rm GeV}$.
The minimal value of Higgs mass correction can be predicted by Eq.~(\ref{mass_corr2}).
For the maximal value of ${\rm Im}[\omega_{ij}]=\omega_{\rm max}$
 (in our numerical analysis $\omega_{\rm max}=\pi$),
 the maximal value of Higgs mass correction is approximately given by the minimal value times $\cosh(2\omega_{\rm max})$.
There is no difference between the NH and the IH cases.
Note that $M_i \gtrsim 10^5\,{\rm GeV}$, which equivalently corresponds to $Y_\nu \gtrsim 1$,
 is excluded by the constraint from lepton flavor violations.\footnote{
The electroweak vacuum becomes instable before the Planck scale
 for ${\rm Tr}[Y_\nu^\dagger Y_\nu] \gtrsim 0.4$~\cite{Rose:2015fua},
 and we have checked that our numerical results shown in Figs.\ref{fig:meffN}-\ref{fig:mue}
 can avoid the vacuum instability except for a few points, at which $M_3 \simeq 10^5\,{\rm GeV}$.
\label{footnote:vacuum}}
If we allow $\mu_i$ to take a larger value than $1\,{\rm MeV}$,
 $Y_\nu$ can be smaller than before,
 and then, there exist allowed regions for $M_i \gtrsim 10^5\,{\rm GeV}$.
However, since such a large $\mu_i$ means a large lepton number violation,
 it conflicts the naturalness.
Thus, we have imposed $\mu_i \leq 1\,{\rm MeV}$.

As we expected in Sec.~\ref{sec:mass_corr},
 the Higgs mass correction can be larger than the Higgs mass for $M_3 \gtrsim 1\,{\rm TeV}$,
 while for $M_3 \gtrsim 10^6\,{\rm GeV}$ in the type-I seesaw model.
This difference corresponds to the difference of size of neutrino Yukawa coupling,
 that is, in the inverse seesaw model $Y_\nu$ is much lager compared with the type-I seesaw case.
The large $Y_\nu$ causes a large mixing between left-handed neutrinos and gauge-singlet neutrinos.
However, such a large mixing can be severely constrained by future experiments of the lepton flavor violation.
In particular, a future experiment of $\mu \to e$ conversion at PRISM can give the strongest constraint.

Figure~\ref{fig:mue} shows the rate of $\mu \to e$ conversion in Titanium (blue dots),
 which has been calculated as in Appendix~\ref{app:mue}.\footnote{
Impact of heavy neutrinos on charged-lepton flavor violation in the inverse seesaw model
 has been also addressed in Ref.~\cite{Abada:2015oba}.}
The cyan dots show the excluded points due to the constraints in Ref.~\cite{Atre:2009rg}.
The black-dashed line corresponds to the future sensitivity of PRISM experiment $R_{\mu \to e} < 10^{-18}$~\cite{Barlow:2011zza},
 while the current upper bound is $R_{\mu \to e} < 4.3\times 10^{-12}$~\cite{Dohmen:1993mp}.
For Titanium,
 the $\mu \to e$ conversion rate vanishes at $M_i \simeq 4.5\,{\rm TeV}$.
Since the vanishing point is the different for the various nuclei,
 the experiment using Titanium can be complemented by experiments using other nuclei.
In addition, the low mass region $M_i \lesssim 140\,{\rm MeV}$
 may be excluded by constraints coming from the big bang nucleosynthesis~\cite{Canetti:2012kh}.
Therefore, we can expect the inverse seesaw model is highly testable,
 i.e., future experiments of neutrinoless double beta decay and $\mu \to e$ conversion
 can search the low mass region $M_i \lesssim 1\,{\rm GeV}$
 and the high mass region $M_i \gtrsim 1\,{\rm GeV}$, respectively.\footnote{
If we allow $\mu_i$ to take a larger value than $1\,{\rm MeV}$,
 there is more allowed region which has a smaller $Y_\nu$.
Actually, the BAU can be explained only in such a parameter space~\cite{Abada:2015rta},
 although it is not preferred from the naturalness point of view.
}

Finally, we mention our assumption ``$M_N$ and $\mu_N=\mu_S$ are diagonal".
If this assumption were relaxed,
 the matrix $V$ is not the identity matrix,
 which can change the configuration of the Yukawa coupling matrix $Y_\nu$ (see Eq.~(\ref{nuYukawa})).
However, this new degree of freedom does not change the order of magnitude of the matrix,
 and the matrix $R$ can also change the configuration of $Y_\nu$.
Thus, we can expect that the effect of $V$ does not change allowed parameter space of $Y_\nu$ (at least significantly).
Actually, in the numerical calculations, the matrix $U$ is rather effective than $Y_\nu$,
 and we have checked that all scatter plots of ($M_i$, $|U_{ij}|^2$) are uniformly distributed,
 which means our assumption ($M_N$ and $\mu_N=\mu_S$ are diagonal) does not restrict the results.
After all, we can expect that, even if the assumption were relaxed, our statement does not change.

%%%%%%%%%%%%%%%%%%%%%%%%%%%%%%%%%%%%%%%%%%%%
\section{Summary} \label{sec:summary}
%%%%%%%%%%%%%%%%%%%%%%%%%%%%%%%%%%%%%%%%%%%%

We focus on the (3, 3) inverse seesaw model,
 in which the number of both right-handed neutrinos and sterile neutrinos are three.
We have investigated heavy neutrino contributions for the neutrinoless double beta decay.
Its rate is proportional to the effective neutrino mass,
 and it is useful to estimate contributions from the heavy neutrinos.
We have found an analytic form of the heavy neutrino contribution to the effective neutrino mass.
It is strongly suppressed for a large heavy-neutrino mass $\gtrsim 1\,{\rm GeV}$,
 while, in $\sim 0.1\,{\rm GeV}$ region, it can be enhanced by ten times or more
 than the active neutrino contribution alone.
We have also investigated the Higgs mass correction coming from the heavy neutrinos,
 and found the minimal value of Higgs mass correction is determined for a given heavy neutrino mass,
 which is usually larger than the type-I seesaw case.
Then, we have shown numerical results of heavy neutrino contributions to 
 the effective neutrino mass and the Higgs mass correction.
As a result, we have found that almost all parameter space of the inverse seesaw model can be complementarily searched:
 the low mass region $M_i \lesssim 1\,{\rm GeV}$ and the high mass region $M_i \gtrsim 1\,{\rm GeV}$
 can be searched by future experiments of neutrinoless double beta decay and $\mu \to e$ conversion, respectively.

%%%%%%%%%%%%%%%%%%%%%%%%%%%%%%%%%%%%%%%%%%%%
\subsection*{\centering Acknowledgment} \label{Acknowledgement}
%%%%%%%%%%%%%%%%%%%%%%%%%%%%%%%%%%%%%%%%%%%%
The authors thank C. Weiland for the important comments about footnote \ref{footnote:one-loop} and \ref{footnote:vacuum}.
This work is partially supported by Scientific Grants
 by the Ministry of Education, Culture, Sports, Science and Technology of Japan
 (Nos. 24540272, 26247038, 15H01037, 16H00871, and 16H02189).
The work of Y. Y. is supported
  by Research Fellowships of the Japan Society for the Promotion of Science for Young Scientists
  (Grants No. 26$\cdot$2428).

%%%%%%%%%%%%%%%%%%%%%%%%%%%%%%%%%%%%%%%%%%%%
\section*{Appendix}
%%%%%%%%%%%%%%%%%%%%%%%%%%%%%%%%%%%%%%%%%%%%
\appendix
%%%%%%%%%%%%%%%%%%%%%%%%%%%%%%%%%%%%%%%%%%%%
\section{$\mu \to e$ conversion rate} \label{app:mue}
%%%%%%%%%%%%%%%%%%%%%%%%%%%%%%%%%%%%%%%%%%%%

Due to the existence of the heavy neutrinos,
 violation of charged lepton number arises at the one loop level.
$\mu \to e$ conversion is induced by a series of gauge boson mediated diagrams.
Its rate is calculated by~\cite{Alonso:2012ji}
 (see also Ref.~\cite{Abada:2014kba} for the inverse seesaw model with or without supersymmetry)
\begin{eqnarray} 
	R_{\mu \to e} \simeq \frac{G_F^2 \alpha_W^2\alpha^3m_\mu^5}{8\pi^4\Gamma_{\rm capt}}\frac{Z_{\rm eff}^4}{Z}F_p^2
	\left| \sum_{i=1}^{9} \left[\left(A+Z\right)F_u(x_i)+\left(2A-Z\right)F_d(x_i)\right] U_{ei}U^*_{\mu i}\right|^2  \,.
\label{muerate}
\end{eqnarray}
 where $\alpha_W=g_2^2/(4\pi)$, $\alpha=e^2/(4\pi)$, $s_W=\sin \theta_W$ is the Weinberg angle,
 $G_F$ is the Fermi constant, and $m_\mu$ is the muon mass.
The other constant parameters depend on a nuclei information which is used in experiments.
$A$ is the mass number, $Z$ ($Z_{\rm eff}$) is the (effective) atomic number,
 $F_p$ is a nuclear form factor, and $\Gamma_{\rm capt}$ is the capture rate.
These values are given in Table~\ref{nuclfactors}~\cite{Alonso:2012ji}.
\begin{table}[t]
\centering
\begin{tabular}{|c|ccc|}
\hline
Nucleus $^A_Z \mbox{N}$ & $Z_{\rm eff}$ & $|F_p(-m^2_\mu)|$ &$ \Gamma_{\rm capt} $ ($10^6s^{-1}$)\\
%\vspace{1mm}
\hline
$_{13}^{27}\mbox{Al}$ & 11.5 & 0.64 & 0.7054\\
$_{22}^{48}\mbox{Ti}$ & 17.6 & 0.54 & 2.59\\
$^{197}_{79} \mbox{Au}$ & 33.5 & 0.16 & 13.07\\
$^{208}_{82} \mbox{Pb}$ & 34.0 & 0.15 & 13.45\\
\hline
\end{tabular} 
\caption{Nuclear form factors and capture rates.}
\label{nuclfactors}
\end{table}

$F_u(x_i)$ and $F_d(x_i)$ are functions of $x_i\equiv m_i^2/M_W^2$ given as
\begin{eqnarray}
	\tilde{F}_u(x)&=&\ \ \frac{2}{3} s^2_W \Big[	F_\gamma(x) -F_Z(x) -2G_Z(0,x)	\Big] \nonumber \\
	&&+ \frac{1}{4} 	\Big[F_Z(x) +2G_Z(0,x)+F_{Box}(0,x)-F_{Box}(0,0)\Big] \,  ,   
\label{functionfutilde} \\ 
	\tilde{F}_d(x)&=&-\frac{1}{3} s^2_W \Big[	F_\gamma(x) -F_Z(x) -2G_Z(0,x)	\Big] \nonumber \\
	&&- \frac{1}{4} 	\Big[F_Z(x) +2G_Z(0,x)-F_{XBox}(0,x)+F_{XBox}(0,0)\Big] \,  ,     
\label{functionfdtilde} \\ 
	F_u(x)&=&\tilde{F}_u(x) +\frac{2}{3} s^2_W G_\gamma(x) \,,
\label{functionfu} \\
	F_d(x)&=& \tilde{F}_d(x) -\frac{1}{3} s^2_W G_\gamma(x) \,.
\label{functionfd}
\end{eqnarray}
The loop functions are 
\begin{eqnarray}
F_\gamma(x)&=& 	\frac{x(7x^2-x-12)}{12(1-x)^3} - \frac{x^2(x^2-10x+12)}{6(1-x)^4} \ln x	\, ,\\
G_\gamma(x)&=&    -\frac{x(2x^2+5x-1)}{4(1-x)^3} - \frac{3x^3}{2(1-x)^4} \ln x \, ,\label{Ggamma} \\
F_Z(x)&=& -\frac{5x}{2(1-x)}-\frac{5x^2}{2(1-x)^2}\ln x \, , \\
G_Z(x,y)&=& -\frac{1}{2(x-y)}\left[	\frac{x^2(1-y)}{1-x}\ln x - \frac{y^2(1-x)}{1-y}\ln y	\right]\, , \\
F_{Box}(x,y)&=&\frac{1}{x-y}\Big\{\left(	4+\frac{xy}{4} 	\right) \left[\frac{1}{1-x}+\frac{x^2}{(1-x)^2} \ln x - \frac{1}{1-y}-\frac{y^2}{(1-y)^2} \ln y\right] \nonumber \\
 	 	&&-2 xy \left[\frac{1}{1-x}+\frac{x }{(1-x)^2} \ln x - \frac{1}{1-y}-\frac{y }{(1-y)^2} \ln y\right]\Big\} \, , \\
F_{XBox}(x,y)&=&\frac{-1}{x-y}\Big\{\left(	1+\frac{xy}{4} 	\right) \left[\frac{1}{1-x}+\frac{x^2}{(1-x)^2} \ln x - \frac{1}{1-y}-\frac{y^2}{(1-y)^2} \ln y\right] \nonumber \\
 	 	&&-2 xy \left[\frac{1}{1-x}+\frac{x }{(1-x)^2} \ln x - \frac{1}{1-y}-\frac{y }{(1-y)^2} \ln y\right]\Big\} \, ,
 \end{eqnarray}
with the limiting values
\begin{align}
G_Z(0,x) & =  -\frac{x}{2(1-x)} \ln x \, ,  \label{limitval1}\\
 F_{Box}(0,x)	 &  =    \frac{4}{ 1-x } + \frac{4x}{(1-x)^2}\ln x\, ,&
F_{XBox}(0,x)  &=   - \frac{1}{ 1-x } - \frac{x}{(1-x)^2}\ln x \, .
\end{align}
As a result, 
 Eqs.~(\ref{functionfu}) and (\ref{functionfd}) are given by
\begin{eqnarray}
	F_u(x) &=& \frac{1}{72(1-x)^4}\Big[ (1-x) x \left\{ 27(1-x)^2 + 4 s_W^2 (31x^2-76x+21) \right\} \nonumber\\
	&& + x \left\{ 27(2-x)(1-x)^2 + 8 s_W^2 (8x^3-11x^2-15x+6) \right\} \ln x \Big]\,, \\
	F_d(x) &=& \frac{1}{72(1-x)^4}\Big[ (1-x) x \left\{ 27(1-x)^2 - 2 s_W^2 (31x^2-76x+21) \right\} \nonumber\\
	&& + x \left\{ 27x(1-x)^2 - 4 s_W^2 (8x^3-11x^2-15x+6) \right\} \ln x \Big]\,.
\end{eqnarray}

%%%%%%%%%%%%%%%%%%%%%%%%%%%%%%%%%%%%%%%%%%%%
%%%%%%%%%%%%%%%%   bibliography   %%%%%%%%%%%%%%%%%%%
%%%%%%%%%%%%%%%%%%%%%%%%%%%%%%%%%%%%%%%%%%%%

\end{document}